\documentclass[page-classic]{epl2}

\title{Quantum effects in spontaneous emission by a relativistic, undulating electron beam}
\shorttitle{Quantum effects in spontaneous emission}
\author{G.R.M. Robb \and R. Bonifacio} 
\institute{ SUPA, Department of Physics, University of Strathclyde,
Glasgow, G4 0NG, Scotland, UK }

\abstract{
Current models of the effect of spontaneous emission on the electron beam dynamics neglect the discreteness of electron recoil associated with photon emission. We present a novel, one-dimensional model of the effect of spontaneous emission on the electron beam dynamics in an undulator both in the classical regime where discrete electron recoil is negligible, and the quantum regime where it is significant. It is shown that in the classical regime, continuous decrease of the average electron energy and diffusive growth of the electron energy spread occurs, in agreement with previous classical models. In the quantum regime, it is shown that the evolution of the electron momentum distribution occurs as discrete momentum groups according to a Poisson distribution. The narrow momentum features of the quantum regime may be useful for generation of coherent radiation, which relies on electron beams having sufficiently narrow momentum/energy distributions.
}

\begin{document}
\maketitle

\section{1. Introduction}

Free electron lasers (FELs) using relativistic electron beams in static undulators or wigglers have recently produced high-intensity coherent radiation in the XUV and X-ray regions of the spectrum \cite{FLASH,LCLS}. Methods for extending the wavelength range of coherent FEL-like sources to even shorter wavelengths using magnetostatic or electromagnetic laser wigglers are being actively investigated \cite{Thomson}. These efforts require us to revisit theoretical models used to describe the emission of radiation by relativistic electrons. In most current models of coherent radiation generation by e.g. the Free Electron Laser (FEL), the discreteness of the recoil due to spontaneously emitted photons can be safely neglected. In previous classical theories \cite{Saldin,Madey} the momentum spread produced by spontaneous emission is described as a continuous diffusive process. At very short wavelengths (high photon momentum/energy) the discreteness of the momentum recoil will eventually become significant. The example we consider is that of a relativistic electron beam passing through a magnetic undulator but the analysis is also applicable to electromagnetic laser wigglers and betatron oscillations in plasma channels \cite{plasma} as in all cases the electrons oscillate transversely and consequently radiate spontaneously. This paper describes a model which allows a description of the effect of spontaneous emission on the electron beam dynamics in both the classical regime where discrete electron recoil is negligible, and the quantum regime where it is significant. In section 2 we describe this model in the limit of weak wiggler fields (wiggler parameter $a_w \ll 1$) where emission occurs only at the fundamental. In section 3 we demonstrate the main features of the classical and quantum regimes of emission, and in section 4 we describe how the model can be extended to include emission at higher harmonics, which are significant for $a_w > 1$.

\section{2. Spontaneous Emission Model}
\label{fundamental}
We use a model based on the evolution of a distribution function for the electron momenta. In this section, we consider spontaneous emission at the fundamental frequency alone, and harmonics are neglected. The effect of higher harmonics is described in section 4. As was shown in \cite{Jackson}, in the limit where $a_w < 1$, an oscillating electron spontaneously emits photons at a rate $R = \frac{2 \pi \alpha a_w^2}{3 \lambda_w}$ per unit distance through the undulator, where $\alpha$ is the fine structure constant, $a_w = \frac{e B_w}{k_w m c}$ is the wiggler parameter and $\lambda_w$ is the undulator/wiggler period. 
As the electron beam passes through the undulator, photons are emitted at random times. Consequently after a distance $z$, different electrons will have emitted different numbers of photons. If the beam is initially monoenergetic there will therefore be a growth of energy/momentum spread i.e. so-called `quantum excitation' of the beam in addition to a decrease in the average beam momentum i.e. `radiation damping'. The novel feature of our model is that it takes account of the fact that spontaneous emission involves emission of photons with momentum $\hbar k$, where $k=2 \pi/\lambda$ is the photon wavenumber and $\lambda$ is the wavelength of the emitted radiation which in the case of the FEL can be expressed as $\lambda= \lambda_w \left( \frac{1 + a_w^2}{2 \gamma^2} \right)$. Consequently the probability of an electron having momentum $p$  will be increased by spontaneous emission from electrons with momentum ($p + \hbar k$) but decreased by spontaneous emission from electrons with momentum $p$. From this physical picture we can write a rate equation for the distribution function due to spontaneous emission alone:
\begin{equation}
\label{finitediff1}
\frac{\partial W(z,p)}{\partial z} = R \;W(z,p+\hbar k) - R\; W(z,p) 
\end{equation}
where $W(p,z) dp$ is the probability of finding an electron between momenta $p$ and $p+dp$.
In order to see the statistical behaviour which this model describes, it is possible to rewrite the electron momentum in terms of number of photons emitted, $N$, where $N$ is a statistical variable. In order to do this, we use the dimensionless momentum variable $p' = p/\hbar k = \frac{p_0}{\hbar k} - N$. Consequently, eq.~(\ref{finitediff1}) can be rewritten as 
\begin{equation}
\label{finitediff2}
\frac{\partial W(z,N)}{\partial z} = R\; W(z,N-1) - R\; W(z,N) 
\end{equation}
which is the master equation for a Poisson process \cite{van Kampen}. This has a solution 
$$
W(z,N) = \frac{<N>^N e^{-<N>}}{N!}
$$
where the mean and variance of $N$ are described by 
$$
\left \langle N \right \rangle = \left \langle \left(N - \left \langle N \right \rangle \right)^2 \right \rangle = Rz
$$.
Consequently the electron momentum $p = p_0 - N \hbar k$ will behave as 
$$
\left \langle p \right \rangle =  p_0 -  \hbar k Rz
$$
$$
\left \langle \left(p - \left \langle p \right \rangle \right)^2 \right \rangle =  \left(\hbar k \right)^2 R z
$$
where for simplicity we have assumed a beam which is initially monoenergetic.
It should be noted that our model neglects the finite linewidth of the spontaneously emitted radiation i.e. it assumes emission of photons of a well-defined momentum $\hbar k$. As the linewidth of wiggler radiation is $\frac{\Delta \omega}{\omega} \sim \frac{1}{N_w}$, then the corresponding uncertainty in photon momentum is $\hbar \Delta k \sim \frac{\hbar k}{N_w}$. Consequently the neglect of the finite linewidth of the radiation in our model is valid if $N_w \gg 1$. For consistency, this means that the probability of photon emission per wiggler period is small i.e. $R \lambda_w \sim 2 \alpha a_w^2 <<1$, which is satisfied if $a_w^2 \ll \frac{1}{2 \alpha} \sim 70$. This assumption is therefore valid for most FELs which use wigglers with $a_w \sim 1$.

\section{3. Classical and Quantum Regimes}
\label{regimes}
In order to observe the different regimes of spontaneous emission, we use the dimensionless variables $P = \frac{p}{\sigma_p}$, $\epsilon = \frac{\hbar k}{\sigma_p}$ and $Z=Rz$, where $\sigma_p$ is a characteristic momentum spread of the system e.g. that produced by interaction or instability such as in the Free Electron Laser (FEL). Note that the characteristic momentum spread, $\sigma_p$, in the case of a low-gain FEL would be that associated with the linewidth of the spontaneous emission curve i.e. $\sigma_p = \sigma_\gamma m c = \frac{\gamma m c}{N_w}$, where $m$ is the electron rest mass, $c$ is the speed of light and $N_w$ is the number of undulator/wiggler periods. In the case of a high-gain FEL, $\sigma_p = \sigma_\gamma m c = \rho \gamma m c$, where $\rho$ is the classical FEL parameter and the parameter $\epsilon$ becomes
$$
\epsilon = \frac{1}{\bar{\rho}}
$$ 
where $\bar{\rho} = \frac{\gamma m c}{\hbar k} \rho$ is the quantum FEL parameter \cite{QFEL1,QFEL2}.

In terms of these variables eq.~\ref{finitediff1} becomes
\begin{equation}
\label{finitediff3}
 \frac{\partial W(Z,P)}{\partial Z} = W(Z,P+\epsilon ) - W(Z,P).
\end{equation}
According to the analysis of eq.~\ref{finitediff2} above, the average value of $P$ will decrease linearly at rate $\epsilon$ i.e. 
\begin{equation}
\label{meanp}
\left \langle P \right \rangle = \left \langle P_0 \right \rangle - \epsilon Z
\end{equation}
and $P$ will diffuse according to
\begin{equation}
\label{diff1}
\sigma^2 \equiv \left \langle (P - \left \langle P \right \rangle)^2 \right \rangle = \epsilon^2 Z
\end{equation}.

\subsection{3.1 The Classical Limit ($\epsilon \rightarrow 0$)}
\label{classicallimit}
In the limit $\epsilon \rightarrow 0$ it is possible to approximate the finite difference equation in eq.~\ref{finitediff2} using a Taylor expansion and retain only the first two terms i.e.
$$
W(Z,P+\epsilon ) \approx W(Z,P) + \epsilon \frac{\partial W(Z,P)}{\partial P} + \frac{\epsilon^2}{2} \frac{\partial^2 W(Z,P)}{\partial P^2} +... 
$$
Consequently, the finite difference equation in eq.~(\ref{finitediff2}) reduces to the drift-diffusion equation 
\begin{equation}
\label{Fokker}
\frac{\partial W(Z,P)}{\partial Z} = \epsilon \frac{\partial W(Z,P)}{\partial P} + \frac{\epsilon^2}{2} \frac{\partial^2 W(Z,P)}{\partial P^2}.
\end{equation}
Converting to unscaled variables, the expression in eq.~\ref{diff1} can be written as 
\begin{equation}
\label{diff2}
\left \langle (\Delta \gamma^2) \right \rangle = \left(\frac{\hbar k}{m c} \right)^2 R z =   \frac{4}{3 \pi^2} k_w^3 r_e \bar{\lambda}_c a_w^2 \gamma^4 z
\end{equation}
which is approximately the same expression (to within a numerical factor $\sim$ 1) as those derived in \cite{Saldin} and \cite{Madey} when $a_w < 1$.

\begin{figure}[htbp]
	\centering
		\includegraphics[width=0.75\columnwidth]{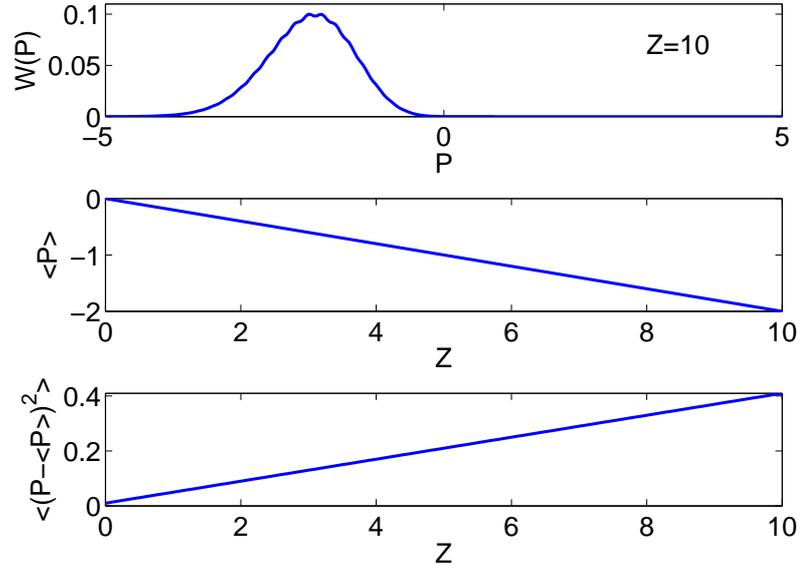}
	\caption{The electron momentum distribution at $Z=10$ and the evolution of the mean and variance of the momentum distribution as calculated from a numerical solution of eq.~\ref{finitediff2} when $\epsilon = 0.2$ and $\sigma_0=0.1$. }
	\label{fig:classical}
\end{figure}

A numerical example of classical evolution is shown in figure~\ref{fig:classical} for the case where $\epsilon = 0.2$. The drift and diffusive growth towards a Gaussian-like momentum distribution can be seen clearly as $Z$ increases.

\subsection{3.2 The Quantum Limit ($\epsilon \geq 1$)}
\label{quantumlimit}
\begin{figure}[htbp]
	\centering
		\includegraphics[width=0.75\columnwidth]{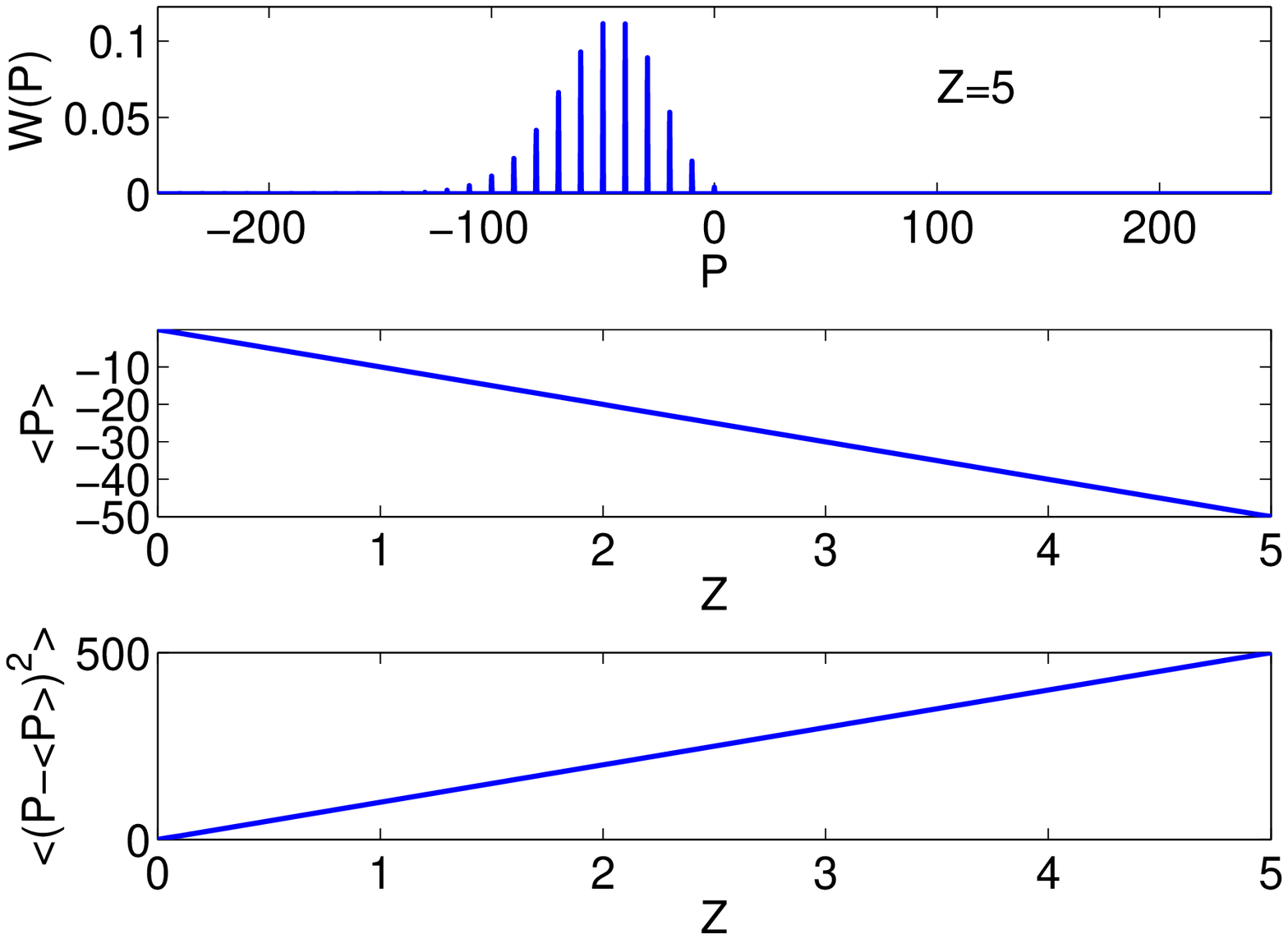}
	\caption{The electron momentum distribution at $Z=5$ and the evolution of the mean and variance of the momentum distribution as calculated from a numerical solution of eq.~\ref{finitediff2} when $\epsilon = 10$ and $\sigma_0=0.1$. }
	\label{fig:quantum}
\end{figure}

In the case where $\epsilon$ is sufficiently large that the higher-order terms in the Taylor expansion of $W(Z,P+\epsilon)$ must be retained, the reduction to a drift-diffusion equation can no longer be performed and the finite difference form of eq.~(\ref{finitediff2}) must be retained. A numerical example of quantum evolution is shown in figure~\ref{fig:quantum} for the case where $\epsilon = 10$ using the same initial condition as that used for fig.~\ref{fig:classical}. It can be seen that in contrast to the classical case discussed previously, the momentum distribution consists of a series of discrete lines separated by $\epsilon$ ($\hbar k$ in unscaled units). It should be noted that while the {\em envelope} of the distribution broadens with increasing $Z$ according to eq.~(\ref{meanp}) and (\ref{diff1}) , the width of the individual discrete momentum groups do not, remaining close to their original values.

\begin{figure}[htbp]
	\centering
		\includegraphics[width=0.75\columnwidth]{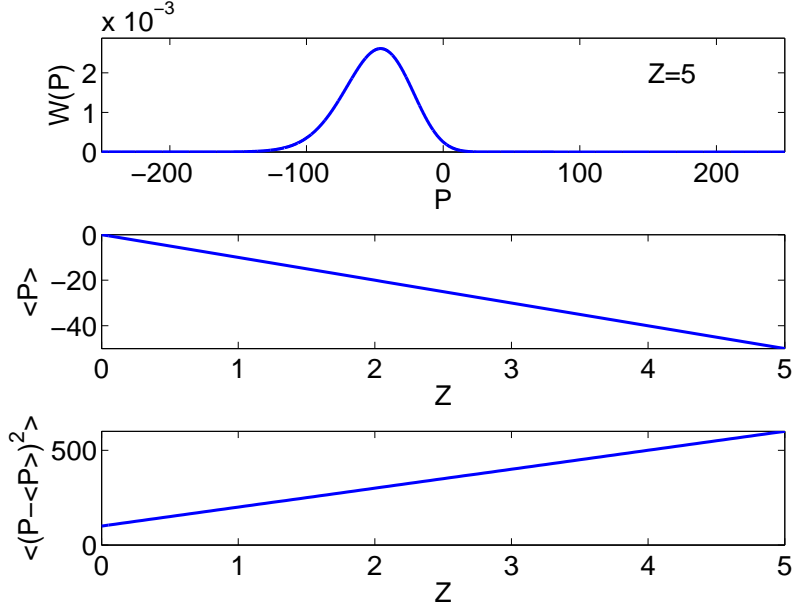}
	\caption{The electron momentum distribution at $Z=5$ and the evolution of the mean and variance of the momentum distribution as calculated from a numerical solution of eq.~\ref{finitediff2} when $\epsilon = 0.2$ and $\sigma_0=10$. }
	\label{fig:spread}
\end{figure}

Note that even for $\epsilon > 1$, in order to observe the quantum discreteness of the momentum distribution, it is necessary that the initial momentum spread , $\sigma_0$, be sufficiently small that the spacing between the momentum lines can be resolved i.e. $\sigma_0 < \epsilon$. This is confirmed in fig.~\ref{fig:spread} which uses the same parameters as in fig.~\ref{fig:quantum} with the exception that the initial spread is now larger i.e. $\sigma_0 = \epsilon = 10 $. It can be seen in fig.~\ref{fig:spread} that the discrete momentum lines visible in ~\ref{fig:quantum} are no longer visible.

\section{4. Spontaneous Emission Including Harmonics}
\label{harmonics}
In the case of a planar undulator where the undulator parameter $a_w > 1$, the emission of radiation at odd harmonics of the fundamental frequency can no longer be neglected. We now extend the model derived in section~\ref{fundamental} to include the effect of radiation at higher harmonics. 

\begin{figure}[htbp]
	\centering
		\includegraphics[width=0.75\columnwidth]{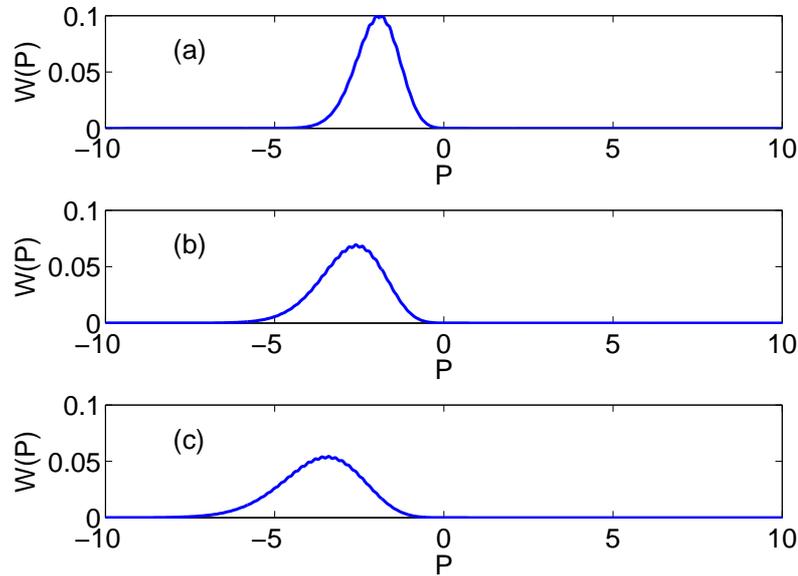}
	\caption{Classical regime including emission at harmonics : Momentum distribution at $Z=10$ calculated from a numerical solution of eq.~(\ref{finitediff_harm2}) including harmonics when $\epsilon = 0.2$ and $\sigma_0=0.1$ for various wiggler parameters $a_w$ : (a) $a_w =0.1$, (b) $a_w=1$ \& (c) $a_w=2$.}
	\label{fig:harm_rhobar5_dist}
\end{figure}

\begin{figure}[htbp]
	\centering
		\includegraphics[width=0.75\columnwidth]{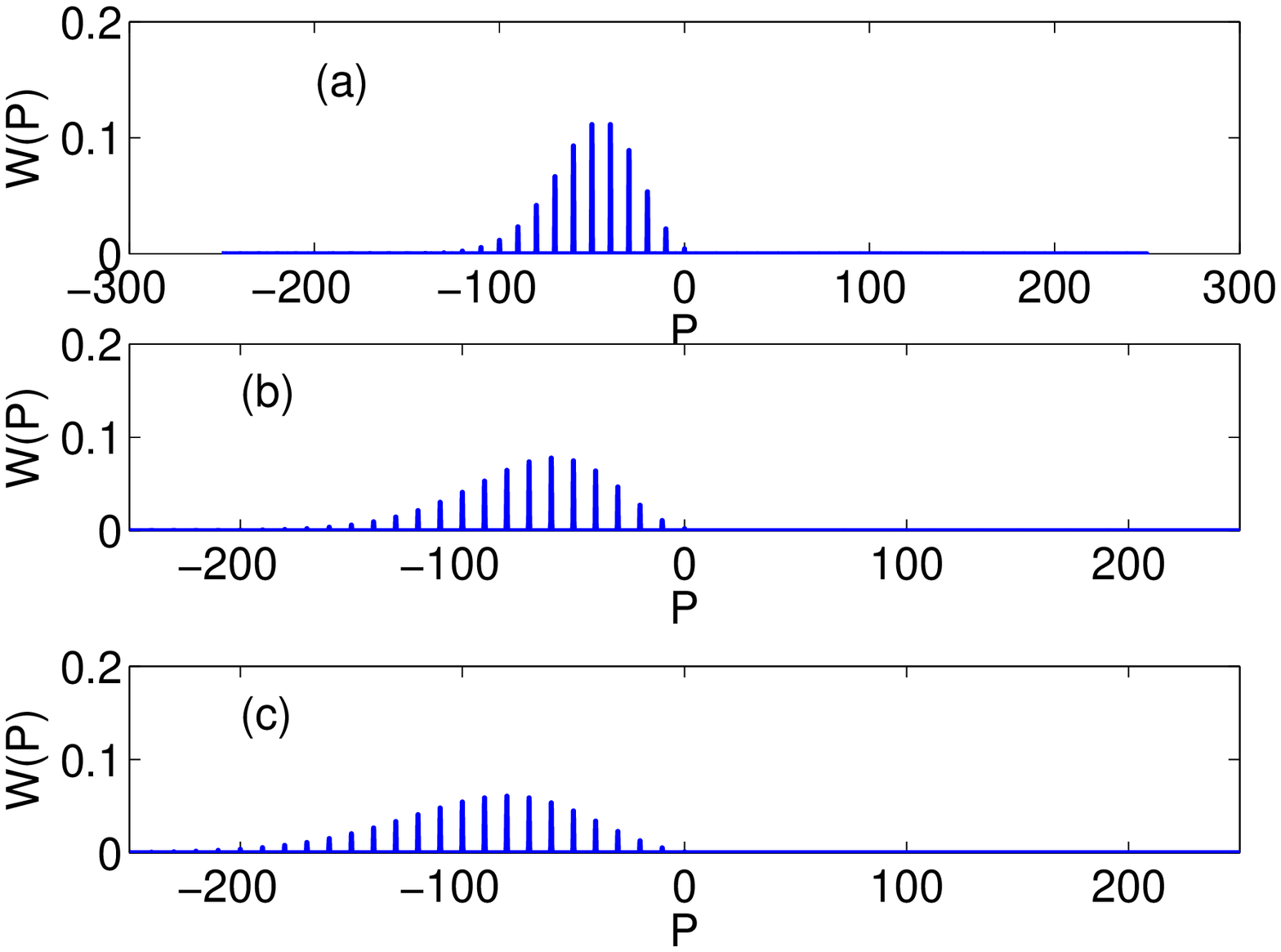}
	\caption{Quantum  regime including emission at harmonics : Momentum distribution at $Z=5$ calculated from a numerical solution of eq.~(\ref{finitediff_harm2}) including harmonics when $\epsilon = 10$ and $\sigma_0=0.1$ for various wiggler parameters $a_w$ : (a) $a_w =0.1$, (b) $a_w=1$ \& (c) $a_w=2$.}
	\label{fig:harm_rhobar01_dist}
\end{figure}

\begin{figure}[htbp]
	\centering
	\includegraphics[width=0.75\columnwidth]{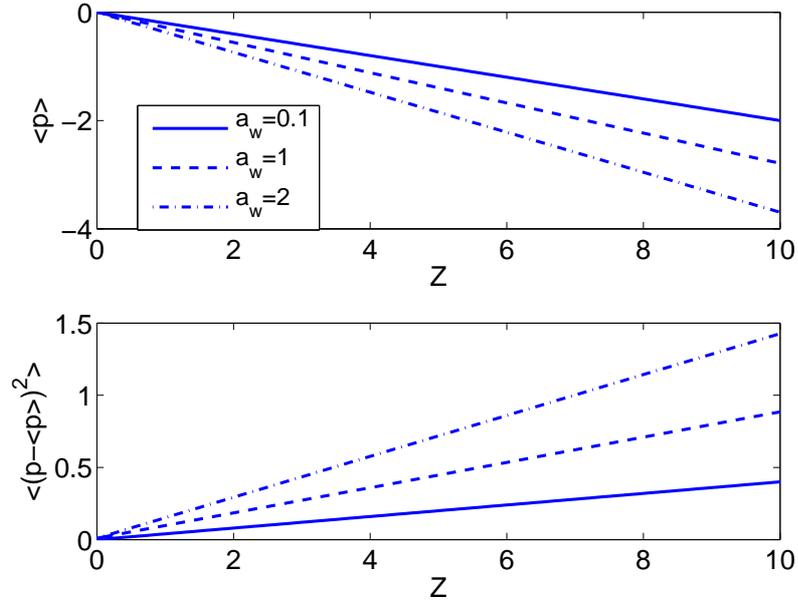}
	\caption{Evolution of mean momentum and momentum variance when $\epsilon = 0.2$ and $\sigma_0=0.1$ (i.e. classical regime) for various wiggler parameters. }
	\label{fig:harm_rhobar5}
\end{figure}

\begin{figure}[htbp]
	\centering
		\includegraphics[width=0.75\columnwidth]{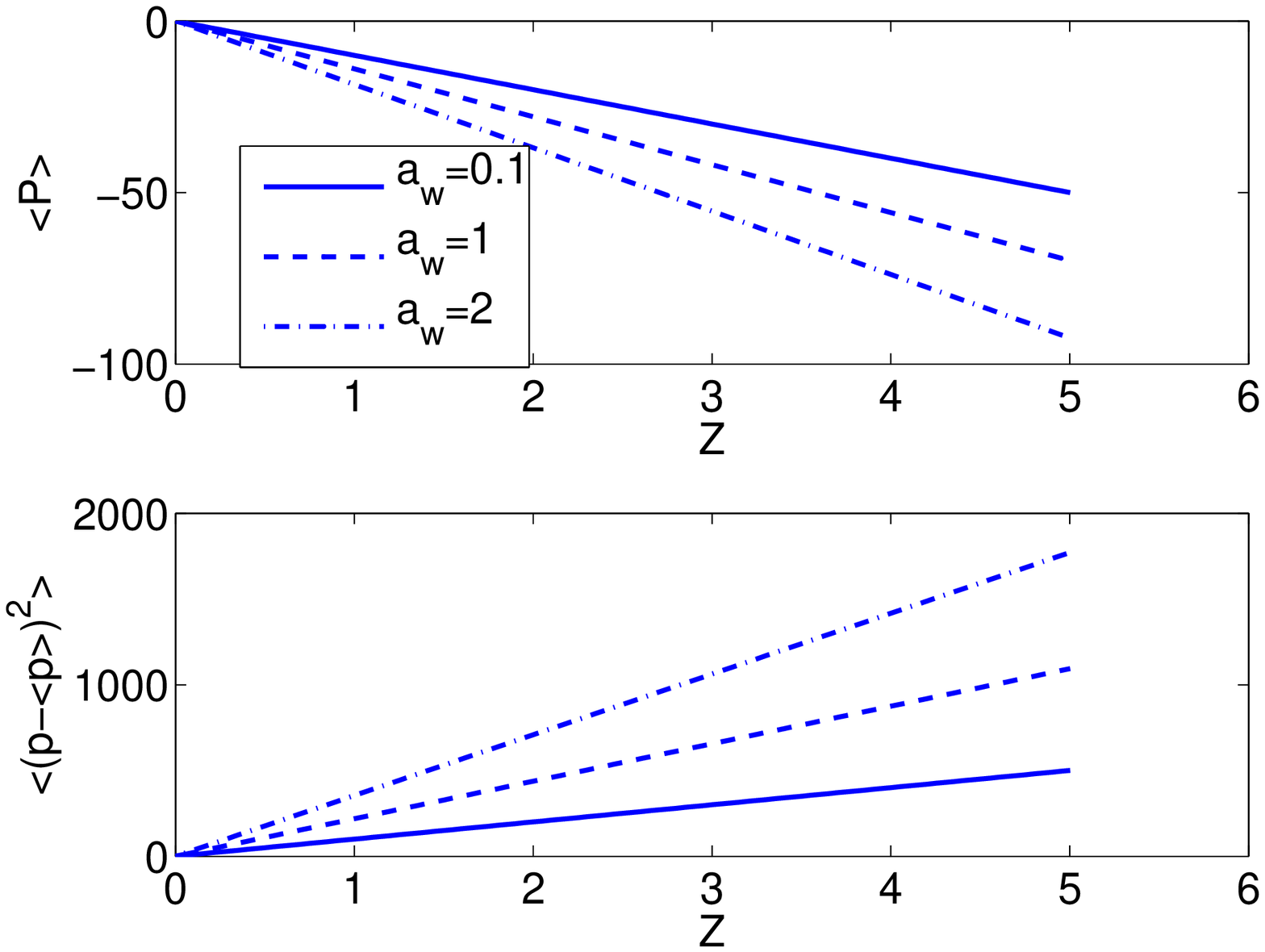}
	\caption{Evolution of mean momentum and momentum variance when $\epsilon = 10$ and $\sigma_0=0.1$ (i.e. quantum regime) for various wiggler parameters. }
	\label{fig:harm_rhobar01}
\end{figure}

When electrons radiate higher harmonics in addition to the fundamental, then the spontaneous emission process involves emission of photons with momentum $n \hbar k$, where $n = 1,3,5,..$. Consequently the probability of an electron having momentum $p$  will be increased by spontaneous emission from electrons with momentum ($p + n \hbar k$) but decreased by spontaneous emission by electrons with momentum $p$. From this physical picture we can write a rate equation for the distribution function due to spontaneous emission including higher harmonics analogous to eq.~(\ref{finitediff1}) :
\begin{eqnarray}
\frac{\partial W(z,p)}{\partial z} 
         &=& \sum_{n=1,3,5,...} \left[ R_n W(z,p+ n \hbar k) - R_n W(z,p) \right] \nonumber \\
         &=& \sum_{n=1,3,5,...} \left[ R_n W(z,p+ n \hbar k) \right] - W(z,p) \left( \sum_{n=1,3,5,...} R_n \right ) \label{finitediff_harm1}
\end{eqnarray}
where $R_n$ is the spontaneous emission rate per unit distance of electrons with momentum $n \hbar k$. $R_1$ is the fundamental spontaneous emission rate and is related to the rate defined in section~\ref{fundamental} by $R_1 = \frac{R}{1 + a_w^2} \left( J_0 (\zeta) - J_1 (\zeta) \right)^2$ where $\zeta = \frac{a_w^2}{2(1+a_w^2)}$ so that $R_1 \rightarrow R$ when $a_w \ll 1$ \cite{Xrayhandbook}.  The emission rate at harmonic $n$ is
\cite{Xrayhandbook}
$$
R_n = \left( \frac{ J_{\frac{n-1}{2}} (n \zeta) - J_{\frac{n+1}{2}} (n \zeta) }{J_0 (\zeta) - J_1 (\zeta)} \right)^2 R_1 .
$$
Repeating the procedure of section~\ref{fundamental} it is possible to rewrite eq.~(\ref{finitediff_harm1}) in terms of the dimensionless variables  $Z = R_1 z$, $P = \frac{p}{\sigma_p}$, and $\epsilon = \frac{\hbar k}{\sigma_p}$, where it should be noted that $Z$ and $P$ have been scaled with respect to the {\em fundamental} emission rate ($R_1$) and photon momentum ($\hbar k$) respectively. Consequently the rate equation including harmonics can be written as 
\begin{eqnarray}
\frac{\partial W(Z,P)}{\partial Z} 
         &=& \sum_{n=1,3,5,...} \left[ \Omega_n W(Z,P+ n \epsilon) - \Omega_n W(Z,P) \right] \nonumber \\
         &=& \sum_{n=1,3,5,...} \left[ \Omega_n W(Z,P+ n \epsilon) \right] - W(Z,P) \left( \sum_{n=1,3,5,...} \Omega_n \right ) \label{finitediff_harm2}
\end{eqnarray}
where 
$$
\Omega_n = \frac{R_n}{R_1} =  \left( \frac{ J_{\frac{n-1}{2}} (n \zeta) - J_{\frac{n+1}{2}} (n \zeta) }{J_0 (\zeta) - J_1 (\zeta)} \right)^2.
$$
Fig.~\ref{fig:harm_rhobar5_dist}  and fig.~\ref{fig:harm_rhobar01_dist} show the influence of harmonics on the evolution of the electron momentum distribution for different wiggler parameters, $a_w$, when $\epsilon = 0.2$ and $\epsilon = 10$ respectively. Fig.~\ref{fig:harm_rhobar5} and fig.~\ref{fig:harm_rhobar01} show corresponding evolution of the mean momentum and the momentum variance for different wiggler parameters, $a_w$, when $\epsilon = 0.2$ and $\epsilon = 10$ respectively. It can be seen from these figures that in both classical and quantum regimes as $a_w$ increases, both the rate of decrease of the mean momentum (drift) and the momentum variance also increase. The reason for this is that when the electrons emit a harmonic, $n$ (where $n > 1$), each spontaneous emission event causes population exchange between momentum groups separated by $n \hbar k$, which are more widely separated in momentum than in the case of emission at the fundamental alone (where n=1).

\section{5. Conclusions}
We have presented a novel dynamical model of spontaneous emission by a relativistic electron beam spontaneously emitting undulator radiation, which is applicable to both magnetostatic wigglers and electromagetic laser wigglers (i.e. Thomson or Compton scattering) in both classical and quantum regimes of emission. Using this model, we have shown that in the classical limit where discrete photon recoil is negligible, the model predicts a linear drift and diffusion of the momentum distribution at a rate which agrees well with those of previous work using alternative classical models \cite{Saldin, Madey}. In addition, our model shows that in the quantum limit where photon recoil is significant the evolution of the momentum distribution changes dramatically to involve discrete narrow momentum groups under an envelope which drifts and diffuses. When emission at harmonics is also significant, it was shown that as $a_w$ increases, both the rate of decrease of the mean momentum (drift) and the momentum variance also increase.

To our knowledge this is the first time that the dynamical evolution of electron beam dynamics due to quantum effects arising from spontaneous emission by a beam of oscillating electrons have been consistently described. By combining this model with the coherent emission processes in e.g. an FEL we should now be able to get a truer picture of the significance of quantum effects associated with spontaneous emission. Of particular interest is its impact on coherent FEL emission at short wavelengths using classical and quantum modes of FEL operation. To date these considerations have been based on the evolution of the {\em envelope} of the momentum distribution, which in the quantum regime can completely neglect the much narrower momentum features associated with discrete photon recoil. As coherent radiation generation typically relies on electron beams with narrow momentum spreads, this difference could be significant. The model described here retains the narrow features of the quantum regime, so consequently should be useful for investigation of schemes in which the aim is control and exploitation of the discrete structure of the electron distribution arising from the quantum regime of spontaneous emission for coherent short-wavelength generation.

\acknowledgments
The authors acknowledge R. Islam, D. Jaroszynski, B.W.J. McNeil \& N. Piovella for helpful discusions. In addition, we acknowledge support from the Leverhulme Trust via research grant research grant F/00273/I, the U.K. EPSRC, the EU FP-7 Laserlab-Europe consortium, and the EU FP7-Extreme Light Infrastructure.

\end{document}